\documentclass[12pt,preprint]{aastex}
\def\paren#1{\left( #1 \right)} 
\def\brace#1{\left\{ #1 \right\}}
\def\bra#1{\left[ #1 \right]}
\def\angle#1{\left\langle #1 \right\rangle}
\begin{document}

\title{Polarized Gravitational Waves from Gamma-Ray Bursts}
\author{Shiho Kobayashi$^{1,2}$ and Peter M\'{e}sz\'{a}ros$^{1,2}$}

\affil{
$^{1}$ Center for Gravitational Wave Physics, Department of Physics,
Pennsylvania State University, University Park, PA 16802\\
$^{2}$Department of Astronomy \& Astrophysics, 
Pennsylvania State University, University Park, PA 16802}

\begin{abstract}

Significant gravitational wave emission is expected from
gamma-ray bursts arising from compact stellar mergers, and 
possibly also from bursts associated with fast-rotating 
massive stellar core collapses. These models have in common 
a high angular rotation rate, and observations provide evidence 
for jet collimation of the photon emission, with properties
depending on the polar angle. Here we consider the gravitational wave
emission and its polarization as a function of angle which is expected 
from such sources. We discuss possible correlations between the 
burst photon luminosity, or the delay between gravitational wave 
bursts and X-ray flashes, and the polarization degree of the 
gravitational waves.

\end{abstract}

\section{Introduction}
\label{sec:intro}

The observations of gamma-ray burst (GRB) afterglows at energies ranging
from X-rays to radio have lead to an increased understanding of the
possible geometry of the inferred relativistic outflow or ejecta 
(see van Paradijs, Kouveliotou \& Wijers 2000 for a review; also
Rhoads 1999; Sari, Piran \& Halpern 1999). At least for the class of
long bursts (durations $\gtrsim$ 10 s) there  is now significant
evidence that the ejecta is collimated in a jet. 

The evidence for jets is based on a change of the GRB light curve
time-dependence power law index, and there appears to be a correlation in  
the sense that bursts with the largest gamma-ray fluences  have the 
narrowest implied opening angles. Frail et al. (2001), Piran et
al. (2001) and Panaitescu \& Kumar (2002) reported that the total 
gamma-ray energy release, after correcting for the collimation and 
distance determined by the afterglow observations,
are narrowly clustered around $5 \times 10^{50}$erg, and suggested that the 
broad distribution in fluence and luminosity for GRBs is largely the result
of a wide variation of the opening angles. 
This result can be obtained assuming that the jet emission inside some
angle $\theta_j$ is approximately uniform and then drops off abruptly
beyond that. However, Rossi, Lazzati \& Rees (2002) and Zhang \&
M\'{e}sz\'{a}ros (2002) showed that the observations can also be
interpreted in terms of an alternative model where the jet, rather than
having a uniform profile out to some definite cone angle, has a
universal angle-dependent beam pattern with a luminosity per unit solid  
angle which is maximal along the axis, and drops off gradually away
from the axis. In this model, it is the difference of viewing angles 
$\theta$ which causes a wide variation in the apparent luminosity of GRBs, 
$L \propto \theta^{-2}$.  

Gravitational waves (GW) are also expected from some types of GRB
(e.g. Kobayashi \& M\'{e}sz\'{a}ros 2002). The GW emissivity and 
its polarization are angle-dependent, and may in principle be measurable, 
depending on the signal to noise ratio and on the detector alignments. 
In this Letter we discuss the
prospects of exploiting measurements of the gravitational polarization
degree to obtain information on the nature and orientation of
the GRB system, providing additional constraints on the luminosity of
the bursts, and we mention possible implications for the interpretation
of dark GRB and X-ray flashes. 

\section{Gravitational Waves from GRB Progenitors}
\label{sec:grb_gw}

Most GRB models involve a stellar system resulting in a rotating black
hole and a massive ($\sim 0.1-1 M_\odot$) disk of accreting matter around it. 
Such massive disks can form from the fall-back of debris during the formation
of the central compact object, which ultimately is likely to lead to a
black hole. Several scenarios could lead to a black hole-accretion 
disk system (e.g. M\'esz\'aros, 2002).  This includes double neutron star 
binary mergers (Eichler et al 1989; Ruffert et al. 1997),
neutron star - black hole mergers (Paczynski 1991;  Janka et al. 1999)
and massive stellar collapses (collapsars) (Woosley 1993; MacFadyen \&
Woosley 1999). 

Coalescing compact binaries are one of the most promising GW sources
detectable by the laser interferometer gravitational wave observatory  
(LIGO), and other detectors such as VIRGO, GEO600 and TAMA300. 
The binary coalescence process can be divided into three phases:
inspiral, merger and ringdown (e.g., Flanagan \& Hughes 1998). 
(1) During the inspiral phase, the gravitational radiation reaction time  
scale is much longer than the orbital period. As the binary losses energy 
by gravitational radiation, the masses gradually spiral in toward each other. 
(2) The merger begins when the orbital evolution is so rapid that
adiabatic evolution is no longer a good approximation. The two masses
then violently merge to form a black hole on a timescale comparable to
a small multiple of the dynamical time.  A bar-mode instability develops 
in the early stages of the merger, releasing a fraction of their rest mass 
energy in GWs. However, a significant fraction of the stellar 
material retains too much angular momentum to cross the black hole horizon 
promptly. This creates a temporary disk of debris material around the black 
hole, whose accretion can power a GRB jet. (3) The black hole
is initially deformed, and in a ringdown phase radiates away the 
energy associated with these deformations as GWs, until it
settles down into a Kerr geometry.   

Massive rotating stellar collapses and their associated GW emission have
been calculated numerically (e.g. Fryer et al. 1999; 
McFadyen \& Woosley 1999; Rampp et al 1998). These numerical estimates are not 
conclusive, as a number of effects (including general relativity, secular  
evolution, etc, see Rampp et al 1998) have been neglected, but they suggest 
that GW emission from massive collapses may be much less 
important than from compact binary mergers. On the other hand, recent 
semi-analytical estimates (van Putten 2001; Fryer, Holz \& Hughes 2002; 
Davies et al 2002) have indicated that massive collapses might emit 
significantly stronger GW signals than expected from the 
previous numerical estimates. Collapsars, i.e. massive stellar collapses 
leading to a GRB, require a high core rotation rate to form a centrifugally 
supported disk around a central, possibly spinning black hole. A high 
rotation rate, however, is conducive to the development of bar or 
fragmentation instabilities in the collapsing core or/and in the massive 
disk (Nakamura  \& Fukugita 1989; Bonnell \& Pringle 1995; van Putten
2002). The asymmetrically infalling matter perturbs the black hole geometry, 
which leads to ringdown gravitational radiation.

\section{Polarization of Gravitational Waves}
\label{sec:gwpol}

We have seen that various mechanisms, including binary, bar and
fragmentation instabilities and oscillations of black holes, could 
occur in GRB progenitor systems, resulting in GW emission. 
It is known that the GWs 
generated by a binary, bar or fragmentation instability are dominated by
modes with spherical harmonic index $l=m=2$ mode (e.g. Misner, Thorne \&
Wheeler 1978). The most slowly damped quasi-normal mode of a Kerr black 
hole also has indices $l=m=2$ (Detweiler 1980). Since this mode may be
preferentially excited in the presence of binary masses or fragmentation
of a massive disk, the ringdown gravitational radiation is also dominated 
by $l=m=2$ mode. 

The polarization components of $l=m=2$ mode depend on the inclination 
angle $\alpha$ (e.g. Thorne 1980) as 
\begin{equation}
h_+ \propto (1+\cos^2\alpha), \ \ \  h_\times \propto 2\cos \alpha.
\end{equation}
GRB progenitors emit GWs more strongly along the polar
axis than in the equatorial plane, the latter being the orbital plane
of the binary, the disk fragments or the bar. Since the GRB jets are 
launched along the polar (angular momentum) axis (i.e. $\alpha=\theta$), 
the GRB progenitors are stronger GW sources than the average non-bursting
merger or collapse, as pointed out by Kochanek and Piran (1993) in the 
context of a neutron star binary model.  When we observe a GRB from a 
viewing angle $\theta$ respect to the polar axis, the 
GW signal is stronger by a factor of 
$(5/32)^{1/2}[(1+\cos\theta)^4+(1-\cos\theta)^4]^{1/2}$ 
than the value averaged over 
all possible viewing angles. Since for the angle dependent jet model 
(Rossi et al. 2002; Zhang \& M\'{e}sz\'{a}ros 2002) the difference of the 
viewing angles causes a variation of the apparent luminosity of the GRB,
we expect a correlation between the apparent luminosity of GRBs and 
the amplitude of the associated GWs, even after distance corrections. 
Also in the case when the jet profile is uniform inside an opening angle 
$\theta_j$, the typical viewing angle is $\sim 2\theta_j/3$, and the 
correlation is still expected. 

The GWs will be circular polarized when viewed along the 
polar axis ($\theta=0$), while the $+$ polarization dominates when viewed along
the equatorial plane ($\theta=\pi/2$). Therefore, the polarization of the
GWs is also correlated with the luminosity of the GRBs. 
We define a polarization tensor in analogy with the electromagnetic approach 
(e.g. Landau \& Lifshitz 1975). A plane wave at the position of a detector 
can be written in the form  
$h_+=\mbox{Re}\brace{A_+ e^{-i\omega t}}$ and 
$h_\times=\mbox{Re}\brace{A_\times e^{-i\omega t}}$.
If the plane wave is monochromatic, the complex amplitudes $A_+$ and
$A_\times$ will be constants, whereas if the wave contains frequencies 
in a small interval $\Delta\omega$ we take $\omega$ to be some average 
frequency inside this range. $A_+$ and $A_\times$ are slowly varying 
functions of time, compared to the wave oscillation period. The 
polarization tensor $\rho_{ab} (a,b=+,\times)$ is defined by using the
time averaged values of the amplitudes $\angle{A_a A_b^\ast}$ and
described by the Stokes parameters $\xi_1$ , $\xi_2$ and $\xi_3$:
\begin{equation}
\rho_{ab}\equiv\frac{\angle{A_a A_b^\ast}}{\angle{|A_+|^2+|A_\times|^2}}
=\frac{1}{2}\paren{ \begin{array}{cc}
            1+\xi_3      & \xi_1-i\xi_2 \\
            \xi_1+i\xi_2 & 1-\xi_3      
           \end{array}}.
\end{equation}
For the $l=m=2$ mode, the parameters are
\begin{eqnarray}
\xi_1&=& 0, \ \ 
\xi_2=\frac{8\cos\theta(1+\cos^2\theta)}
{(1+\cos\theta)^4+(1-\cos\theta)^4}, \\ 
\xi_3 &\equiv& P= \frac{2(1+\cos\theta)^2(1-\cos\theta)^2}
{(1+\cos\theta)^4+(1-\cos\theta)^4}.
\end{eqnarray}
The degree of circular polarization and of linear polarization are given
by  $\xi_2$ and  
$\sqrt{\xi_1^2+\xi_3^2}=\xi_3\equiv P$, respectively. Thus, GRB progenitors
whose polar axes are directed towards the earth would produce  brighter 
GRBs, and circularly polarized GWs. As the viewing angle 
becomes larger relative to the polar axis, the luminosity of the GRB 
decreases $\propto \theta^{-2}$, while the degree of linear polarization
increases as $P \sim 10^{-2} (\theta/30 ~\mbox{degree})^4$.

\section{Polarization Measurement}
\label{sec:polmes}

The response of an interferometer (interferometer $1$) 
to the gravitational radiation is given by a linear combination of 
two polarization components $m_1=F_{+,1}h_++F_{\times,1} h_\times$
where the antenna patterns $F_{+,1}$ and $F_{\times,1}$ depend on the 
orientation of the interferometer with respect to the GW
source (e.g. Finn \& Chernoff 1993). 
We assume that the interferometer arms are of the same
length and that they meet at right angles. Define a right-handed
coordinate system with one interferometer arm along the $x$-axis
and the other along the $y$-axis. For simplicity, we assume that the
source is in the direction of the $z$-axis. Since we can determine the
position of a source in the sky by using observations of the GRB and 
afterglow, it is possible to generalize the following discussion to the 
case of a source with an arbitrary sky position. The angular momentum
vector of the source (direction of the GRB jet) may be oriented in an 
arbitrary direction. We assume that the projection of the angular momentum 
vector to the $x-y$ plane makes an angle $\zeta$ with the $x$-axis. 
With these conventions, the antenna patterns are given by  
$F_{+,1}=\cos 2\zeta$ and $F_{\times,1}=\sin 2\zeta$.

To determine the polarization of GWs, one needs a network consisting of
at least two interferometers which have different (i.e. 
non-parallel) arm orientations. Consider an identical interferometer 
(interferometer 2) at the same location as interferometer 1. 
(If we set the arrival time of the GRB 
signal as the origin of time at each interferometer, we can correct for the actual
physical separation so that the two interferometers can be always considered to be
at the same location). We assume that the  interferometer 2 is rotated by an angle 
$-\pi/4$ around the $z$-axis with respect to the interferometer 1. The response 
of the interferometer 2 is $m_2=F_{+,2}h_++F_{\times,2} h_\times$, and
the antenna patters are given by 
$F_{+,2}=-\sin 2\zeta,$ and $F_{\times,2}=\cos 2\zeta$.

The detection of the polarization of GWs requires
observations with a high signal-to-noise ratio (SNR) $\rho$. A detection 
is likelier in an optimal case where the wave forms of the polarized 
components, $f_{+}$ and $f_{\times}$, are known.
We define the noise-weighted inner product as
\begin{equation}
\angle{s_i, f_a}=4~\mbox{Re}\int_0^\infty
\frac{\tilde{s_i}(f)\tilde{f_a^\ast}(f)}{S_h(f)}df
\end{equation}
where $\tilde{s_i}(f)$ denotes the Fourier transform of the 
outputs of the two  interferometers $s_i(t)=m_i(t)+n_i(t)$ $(i=1,2)$, $n_i(t)$
is the noise of the interferometers, $S_h(f)$ is the one-sided noise power
spectral density and $\tilde{f_a}$ is the Fourier transform of $f_a$
$(a=+,\times)$. We normalize the functions $f_a$ as 
$\angle{f_a,f_b}=\delta_{ab}$, hence the dispersion of the noise
is unity $\sigma^2=\overline{\angle{n_i,f_a}^2}=1$ where the overline
represents ensemble average.
The linear polarization degree $P=\xi_3$ is 
\begin{equation}
P=\frac{\angle{m_1,f_+}^2+\angle{m_2,f_+}^2
           -\angle{m_1,f_\times}^2-\angle{m_2,f_\times}^2}
{\angle{m_1,f_+}^2+\angle{m_2,f_+}^2
           +\angle{m_1,f_\times}^2+\angle{m_2,f_\times}^2}.
\label{eq:P}
\end{equation}

Since the outputs of the interferometers $s_i$ are always superpositions of
signals $m_i$ and noises $n_i$, the inner products $\angle{m_i,f_a}$
themselves are not observable. We estimate $\angle{m_i,f_a}$ by   
$\angle{m_i,f_a}\sim \angle{s_i,f_a}$. Substituting this 
into eq. (\ref{eq:P}), we can get an estimate $P^\prime$ of the
polarization degree. This approximation introduces an error $\Delta$ in
the estimate of the polarization degree $P$, given by
\begin{eqnarray}
\Delta&\equiv&\frac{P^\prime-P}{P}
=\frac{1+(A-B)/P}{1+A+B} -1,\\
A&=&2a_+(\lambda_{1+}\cos2\zeta-\lambda_{2+}\sin2\zeta)+
\lambda_{1+}^2+\lambda_{2+}^2, \\
B&=& 2a_\times(\lambda_{1\times}\sin2\zeta+\lambda_{2\times}\cos2\zeta)
+\lambda_{1\times}^2+\lambda_{2\times}^2,
\end{eqnarray}
where 
$a_+=\sqrt{2}(1+\cos^2\theta)[(1+\cos\theta)^4+(1-\cos\theta)^4]^{-1/2}$,
$a_\times=2\sqrt{2}\cos\theta[(1+\cos\theta)^4+(1-\cos\theta)^4]^{-1/2}$,
$\lambda_{ia}=\angle{n_i,f_a}/\rho$ and 
the SNR $\rho$ is $\rho^2=\angle{h_+,f_+}^2+\angle{h_\times,f_+}^2$.
The fluctuations $\lambda_{ia}$ are of order of $\rho^{-1}$. Thus,
a SNR of $\rho > P^{-1}$ is required, if one wants 
to determine the polarization degree to an accuracy of order $P$.

\begin{figure}
\plotone{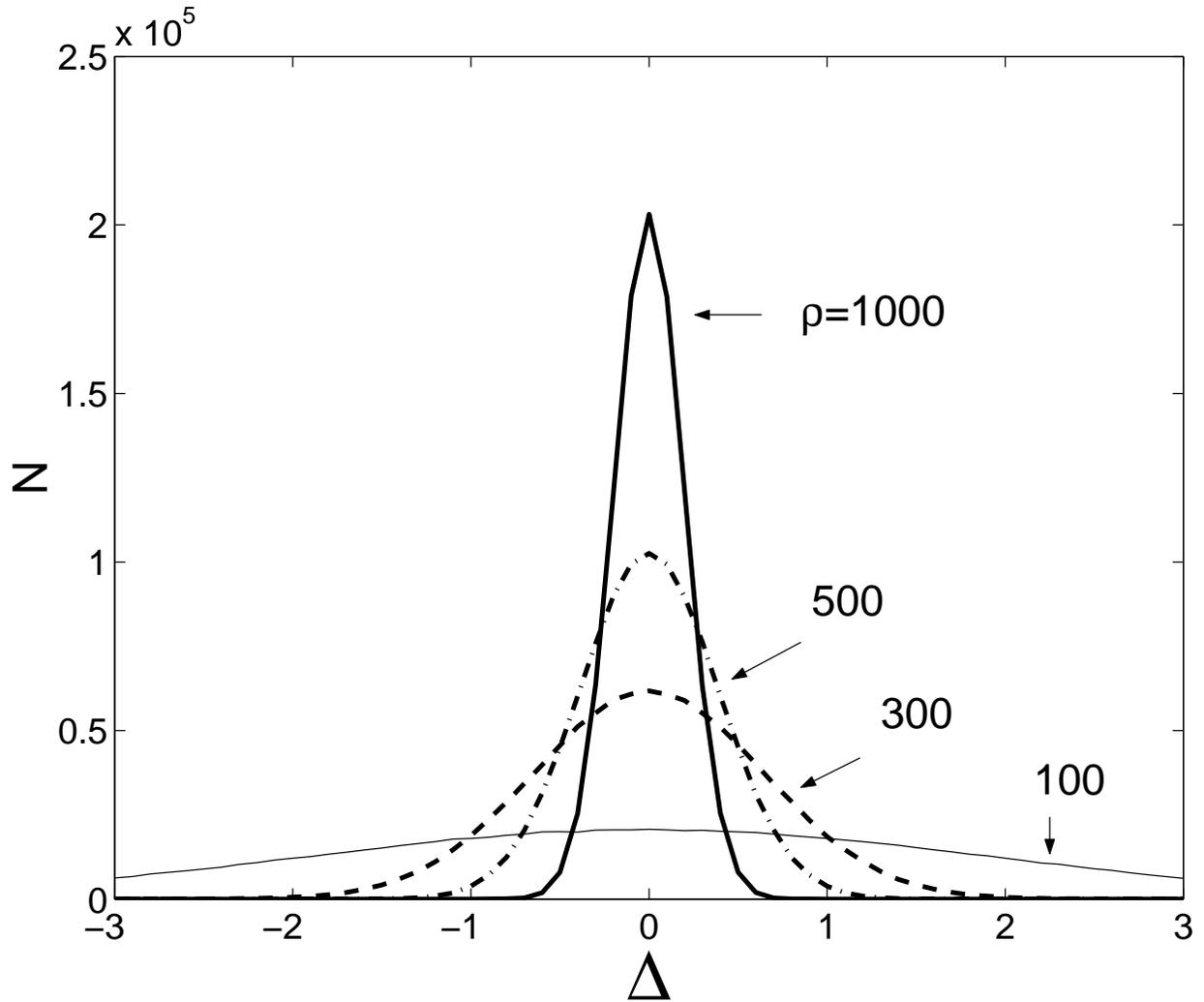}
\caption{Distribution of errors $\Delta$ for a polarization degree 
$P=10^{-2}$, signal to noise ratios $\rho=100$ (thin solid), 300 (dashed), 
500 (dashed dotted) and 1000 (thick solid), based on $10^6$ random 
realizations. The size of the bins is 0.1.
\label{fig1}}
 \end{figure}

To evaluate the error $\Delta$ numerically, we assume that the 
$\lambda_{ia}$ are normally distributed, and that the projection of the 
angular momentum on the orbital plane $\zeta$ relative to the $x$-axis  
is uniformly distributed between $0$ and $2\pi$. 
Figure \ref{fig1} shows the distributions of $\Delta$ for $10^6$ random
realizations in the case of a $1\%$ polarization degree
($P=10^{-2}$). The SNR was assumed to be $\rho=100, 300, 500$ or
1000, for which the fractions of the realizations with 
large errors ($|\Delta|>0.5$) are  80$\%$, 44$\%$, 20$\%$ and 1$\%$,
respectively. Therefore, $\rho\sim10 P^{-1}=1000$ is required to
determine a $1\%$ polarization degree.

The LIGO interferometers are coaligned, and thus cannot by 
themselves determine the polarization degree of GWs, 
although a network consisting of LIGO and other interferometers could 
in principle do it. We assume here an interferometer identical to the 
LIGO detectors with an optimal orientation (rotated by $\pi/4$ with 
respect to the LIGO interferometers). If some fraction of GRB progenitors 
are double neutron star mergers, the SNR is given by 
$\rho_{ave} \sim 16 (d/100 \mbox{Mpc})^{-1}$ (e.g. eq. (16) in 
Kobayashi \& M\'{e}sz\'{a}ros 2002) where $d$ is the distance to the
binary. This estimate was obtained from an average over different 
possible orientation of the source and interferometer. The SNR $\rho$ is
larger by factor of 
$(5/4)\bra{(1+\cos\theta)^4+(1-\cos\theta)^4}^{1/2}$. Then, 
in an optimal case, we can determine a $1\%$ polarization degree up to 
$d_{max} \sim 7$ Mpc. If we assume that $\sim 1000$ GRBs happen in a year
within $\sim 3000$Mpc, the closest event in a year is, on average, at 
$d \sim 300$ Mpc. Since the position in the sky is random, the SNR is 
smaller by a factor of $\sqrt{2/5}$ than in the direction of the z-axis. 
The LIGO is most  sensitive around $f_0\sim 150$Hz and the sensitivity
is about $\sqrt{f_0 S_h(f_0)} \sim 3\times10^{-23}$. Since
$d_{max}$ is proportional to $S_h^{-1/2}$,  a future detector with
sensitivity  $\sqrt{f_0 S_h(f_0)} \sim 4\times10^{-25}$  would enable us 
to measure a polarization degree of $1\%$ in a timescale of one year.

Frail et al. (2001) show that the observed distribution of jet opening  
angles (or viewing angles for the angle dependent jet model) is
$f_{obs} \propto \theta_j^{-3.5}$ ($\theta_j>0.05$). Assuming this
result, the probability that GWs from a GRB have a
polarization degree larger than $P=10^{-2}$ is given by $\sim
3\times 10^{-2} (P/10^{-2})^{-0.4}$. However, the $f_{obs}$ are 
estimated from observations of typical GRBs at redshifts $z\sim 1$,
and one might fail to detect a significant fraction of the dim GRBs at
large viewing angles. Therefore, when we study GWs from 
nearby GRBs, the probability that one observes GWs with 
a large polarization degree could be much larger than the above estimate.

\section{Discussion and Conclusions}
\label{sec:disc}

We have argued that GRB progenitors are likely to emit $l=m=2$ gravitational
waves (GWs), which are circularly polarized on the polar axis, while the $+$
polarization dominates on the equatorial plane. Recent GRB studies suggest 
that the wide variation in the apparent luminosity of GRBs are caused by 
differences in the viewing angle, or possibly also in the jet opening angle. 
Since GRB jets are launched along the polar axis of GRB progenitors, 
correlations among the apparent luminosity of GRBs and the amplitude as 
well as the polarization degree of the GWs are expected.

At a viewing angle larger than the jet opening angle $\theta_j$ (which
may be defined also in the case of a universal angle-dependent jet profile)
the GRB $\gamma$-ray emission may not be detected. However, in such cases
an ``orphan'' long-wavelength afterglow could be observed, which would be
preceded by a pulse of GWs with a significant linearly
polarized component. An expanding jet with an opening angle $\theta_j$ 
behaves, as long as its Lorentz factor  $\gamma > \theta_j^{-1}$, as if it 
were part of a spherical shell, but relativistic beaming effect allows only 
observers at viewing angles $< \theta_j$ to observe the emission from the
jet. (By contrast, since GW emission is expected from the central engine
itself, GWs are not subject to such extreme beaming). As the jet slows down and 
reaches $\gamma \sim \theta_j^{-1}$, the jet begins to expand laterally, and 
its electromagnetic radiation begins to be observable over increasingly 
wider viewing angles. Since the opening angle increases as $\sim\gamma^{-1} 
\propto t^{1/2}$ (Sari et al 1999), at a viewing angle $\theta > \theta_j$, 
the orphan afterglow begins to be observed (or peaks) at a time $t_{p}
\propto \theta^2$ after the detection of the GW burst.
The polarization degree and the peak time should be correlated as
$P\propto t_p^2$. 

A new type of fast transient source, called ``X-ray flashes'', have
recently been observed with the BeppoSAX satellite (Kippen et al. 2002). 
Apart from their large fraction of X-rays ($\sim 2-10$ keV), the overall
properties of these events are similar to those of GRBs. Recently,
Yamazaki, Ioka and Nakamura (2002) and Woosley, Zhang \& Heger (2002)
suggested that these events may be GRBs with large viewing angles. If
this is the case, linearly polarized  
GWs should be observed prior to the X-ray flashes. The
degree of polarization should be positively correlated with longer delays 
and with the softness of the X-ray flashes, which increase with angle.

Since the degrees of linear and circular polarization depend on the 
viewing angle, a determination of the polarization degree would be a
measure of the viewing angle. Such measurements, which are likely to 
require the advent of a future generation of detectors, could provide 
a new tool for estimating the absolute luminosity of GRBs, including
its photon component. By comparing the estimated absolute photon 
luminosity with the apparent luminosity, the distance to the source may 
be estimated independently of any redshift measurement. No optical 
afterglows have been found for about half of all the GRBs detected by 
BeppoSAX (the so called ``dark GRBs''), and the present method would 
have the potential to help determine or constrain the distances to such 
dark GRBs.

We thank L. S. Finn, B. Owen, I. Jones and P. Sutton for useful 
discussions. We acknowledge support through the Center for Gravitational
Wave Physics, which is funded by NSF under cooperative agreement PHY
01-14375, and through NSF AST0098416 and NASA NAG5-9192.

\noindent {\bf References}\newline
Bonnell, I.A. \& Pringle, J.E. 1995, MNRAS, 273, L12.\newline
Davies, M.B., King, A.,Rosswog,S. \& Wynn,G. 2002, ApJ, 579, L63.\newline
Detweiler,S. 1980, ApJ, 239, 292.\newline
Eichler, D., Livio,M., Piran,T. \& Schramm,D.N. 1989, Nature, 340, 126.\newline
Finn, L.S. \& Chernoff,D.F. 1993, Phys. Rev. D, 47, 2198.\newline
Flanagan,E.E. \& Hughes, S.A. 1998, Phys. Rev. D, 57, 4535.\newline
Frail,D.A. et al. 2001, ApJ, 562, L55.\newline
Fryer, C.L., Woosley,S.E., Herant,M. \& Davies,M.B. 1999, ApJ, 520, 650.\newline
Fryer, C.L., Holz, D.E. \& Hughes,S.A. 2002, ApJ, 565, 430.\newline
Janka,H.T., Eberl,T., Ruffert,M. \&  Fryer,C. 1999, ApJ, 527, L39.\newline
Kippen, R.M., Woods,P.M., Heise,J., in'tZand,J.J.M., Briggs,M.S. \&
Preece,R.D. 2002, preprint (astro-ph/0203114).\newline
Kobayashi,S \& M\'{e}sz\'{a}ros,P. 2002, submitted to ApJ,
astro-ph/0210211.\newline
Kochanek,C. \& Piran, T. 1993, ApJ, 417, L17.\newline
Landau,L.D. \& Lifshitz,E.M. 1975, The Classical Theory of Fields,
(New York, Pergamon).\newline
MacFadyen,A.I. \& Woosley, S.E. 1999, ApJ, 524, 262.\newline
M\'{e}sz\'{a}ros, P. 2002, Annu. Rev. Astron. Astrophys., 40,137.\newline
Misner,C.W., Thorne,K.S. \& Wheeler,J.A., Gravitation (Freeman, 
San Francisco 1973).\newline
Nakamura,T. \& Fukugita,M. 1989, ApJ, 337, 466.\newline
Rampp,M., Muller,E. \& Ruffert,M. 1998, A\&A, 332, 969.\newline
Rossi,E., Lazzati,D. \& Rees,M.J. 2002, MNRAS, 332, 945.\newline
Paczynski,B. 1991, Acta Astronomica, 41, 257.\newline
Panaitescu,P. \& Kumar,P. 2002, ApJ, 571, 779.\newline
Piran,T., Kumar,P., Panaitescu,A. \& Piro,L. 2001, ApJ, 560, L167.\newline
Rhoads,J.E. 1999, ApJ, 525, 737.\newline
Ruffert,M., Janka,H.T., Takahashi,K. \& Schaefer,G  1997, A\&A, 319, 122.\newline
Sari,R., Piran,T.\& Halpern,J.P. 1999, ApJ, 519, L17.\newline
Thorne,K.S. 1980, Rev. Mod. Phys., 52, 299.\newline
Yamazaki, R., Ioka,K. \& Nakamura,T. 2002, ApJ, 571, L31.\newline
van Paradijs,J., Kouveliotou,C. \& Wijers,R.A.M.J. 2000, 
Annu. Rev. Astron. Astrophys., 38,379.\newline
van Putten, M.H.P.M. 2001 , ApJ Lett., 562, L51.\newline
van Putten, M.H.P.M. 2002 , ApJ Lett., 575, L71.\newline
Woosley,S. 1993, ApJ, 405, 273.\newline
Woosley,S. Zhang,W \& Heger,A. 2002 , preprint (astro-ph/0206004).\newline
Zhang,B. \& M\'{e}sz\'{a}ros,P. 2002, ApJ, 571, 876.\newline
\end{document}